\documentstyle[twocolumn,aps,epsfig]{revtex}

\catcode`@=11
\def\citer{\@ifnextchar [{\@tempswatrue\@citexr}{\@tempswafalse\@citexr[]}}
 
\def\@citexr[#1]#2{\if@filesw\immediate\write\@auxout{\string\citation{#2}}\fi
  \def\@citea{}\@cite{\@for\@citeb:=#2\do
    {\@citea\def\@citea{--\penalty\@m}\@ifundefined
       {b@\@citeb}{{\bf ?}\@warning
       {Citation `\@citeb' on page \thepage \space undefined}}%
\hbox{\csname b@\@citeb\endcsname}}}{#1}}
\catcode`@=12

\def\refeq#1{\mbox{eq.~(\ref{#1})}}
\def\refeqs#1{\mbox{eqs.~(\ref{#1})}}
\def\reffi#1{\mbox{Fig.~\ref{#1}}}

\def\citere#1{\mbox{Ref.~\cite{#1}}}
\def\citeres#1{\mbox{Refs.~\cite{#1}}}

\newcommand{\mst}{m_{\tilde{t}}}

\newcommand{\mste}{m_{\tilde{t}_1}}
\newcommand{\mstz}{m_{\tilde{t}_2}}

\newcommand{\MstL}{M_{\tilde{t}_L}}
\newcommand{\MstR}{M_{\tilde{t}_R}}

\newcommand{\Mtlr}{M_{t}^{LR}}

\newcommand{\msq}{m_{\tilde{q}}}

\newcommand{\sfl}{\tilde{f}_L}
\newcommand{\sfr}{\tilde{f}_R}

\newcommand{\Pe}{\phi_1}
\newcommand{\Pz}{\phi_2}
\newcommand{\PePz}{\phi_1\phi_2}
\newcommand{\mpe}{m_{\Pe}}
\newcommand{\mpz}{m_{\Pz}}
\newcommand{\mpez}{m_{\PePz}}

\newcommand{\oaas}{{\cal O}(\alpha\alpha_s)}
\newcommand{\cp}{{\cal CP}}

\newcommand{\edz}{\frac{1}{2}}
\newcommand{\twol}{two-loop}
\newcommand{\onel}{one-loop}

\newcommand{\tc}{{\em TwoCalc}}
\newcommand{\fa}{{\em FeynArts}}

\newcommand{\MW}{M_W}
\newcommand{\MZ}{M_Z}
\newcommand{\MA}{M_A}
\newcommand{\mh}{m_h}

\newcommand{\mt}{m_{t}}

\newcommand{\mgl}{m_{\tilde{g}}}

\newcommand{\Stop}{\tilde{t}}
\newcommand{\StopL}{\tilde{t}_L}
\newcommand{\StopR}{\tilde{t}_R}
\newcommand{\Stope}{\tilde{t}_1}
\newcommand{\Stopz}{\tilde{t}_2}

\newcommand{\tst}{\theta_{\tilde{t}}}

\newcommand{\tsf}{\theta\kern-.20em_{\tilde{f}}}
\newcommand{\tsfp}{\theta\kern-.20em_{\tilde{f}\prime}}
\newcommand{\tsq}{\theta\kern-.15em_{\tilde{q}}}
\newcommand{\sw}{s_W}

\newcommand{\KL}{\left(}
\newcommand{\KR}{\right)}

\newcommand{\VL}{\left( \begin{array}{c}}
\newcommand{\VR}{\end{array} \right)}
\newcommand{\ML}{\left( \begin{array}{cc}}
\newcommand{\MLd}{\left( \begin{array}{ccc}}
\newcommand{\MLv}{\left( \begin{array}{cccc}}
\newcommand{\MR}{\end{array} \right)}

\newcommand{\hc}{\mbox {h.c.}}
\newcommand{\re}{\mbox {Re}\,}

\newcommand{\Tb}{\tan \beta\hspace{1mm}}

\newcommand{\CTb}{\cot \beta\hspace{1mm}}

\newcommand{\Sb}{\sin \beta\hspace{1mm}}
\newcommand{\SQb}{\sin^2\beta\hspace{1mm}}
\newcommand{\SDb}{\sin^3\beta\hspace{1mm}}
\newcommand{\Cb}{\cos \beta\hspace{1mm}}
\newcommand{\CQb}{\cos^2\beta\hspace{1mm}}
\newcommand{\CDb}{\cos^3\beta\hspace{1mm}}

\newcommand{\tev}{\,\, {\mathrm TeV}}
\newcommand{\gev}{\,\, {\mathrm GeV}}

\newcommand{\BC}{\begin{center}}
\newcommand{\EC}{\end{center}}
\newcommand{\BE}{\begin{equation}}
\newcommand{\EE}{\end{equation}}
\newcommand{\BEA}{\begin{eqnarray}}
\newcommand{\BEAnn}{\begin{eqnarray*}}
\newcommand{\EEA}{\end{eqnarray}}
\newcommand{\EEAnn}{\end{eqnarray*}}
\newcommand{\non}{\nonumber}
\newcommand{\id}{{\rm 1\kern-.12em
\rule{0.3pt}{1.5ex}\raisebox{0.0ex}{\rule{0.1em}{0.3pt}}}}

\def\al{\alpha}
\def\als{\alpha_s}

\def\be{\beta}

\def\de{\delta}

\def\la{\lambda}

\def\Si{\Sigma}

\def\Sie{\Sigma^{(1)}}
\def\Siz{\Sigma^{(2)}}

\def\hSi{\hat{\Sigma}}

\def\hSiz{\hat{\Sigma}^{(2)}}

\begin{document}                                                              

\title{
{\normalsize
\hfill KA-TP-2-1998\\
\hfill hep-ph/9803277\\[.1em]}
QCD Corrections to the Masses of the neutral
$\cp$-even Higgs Bosons in the MSSM}

\author{ S.\ Heinemeyer, W.\ Hollik and G.\ Weiglein \\}
\address{
Institut f\"ur Theoretische Physik, Universit\"at Karlsruhe,
D--76128 Karlsruhe, Germany \\}
\date{\today}
\maketitle
\begin{abstract}
We perform a diagrammatic calculation of
the leading two-loop QCD corrections to the masses of the
neutral $\cp$-even
Higgs bosons in the Minimal Supersymmetric Standard Model (MSSM).
The results are valid for arbitrary values of the parameters of the
Higgs and scalar top sector of the MSSM.
The \twol\ corrections are found to reduce the mass of the lightest
Higgs boson considerably compared to its one-loop value.
The numerical results are analyzed in the GUT favored 
regions of small and large $\Tb$. 
Their impact on a precise prediction for the mass of the lightest Higgs
boson is briefly discussed.
\end{abstract}


\begin{narrowtext}

Supersymmetric theories (SUSY)~\cite{mssm} are the best motivated
extensions of the Standard Model (SM) of the electroweak and strong
interactions. They provide an elegant way to break the electroweak
symmetry and to stabilize the huge hierarchy between the GUT and the
Fermi scales, and allow for a consistent unification of the gauge
coupling constants as well as a natural solution of the Dark Matter
problem; for recent reviews see Ref.~\cite{reviews}. 
The MSSM predicts the existence of scalar partners $\sfl, \sfr$ to
each SM chiral fermion, and spin-$1/2$ partners to the gauge bosons
and to the scalar Higgs bosons. So far, the direct search for SUSY
particles has not been successful. One can only set lower bounds of
${\cal O}(100) \gev$ on their masses~\cite{PDG}.

A particularly stringent test of the MSSM is the
search for the lightest Higgs boson. At tree level its mass, $\mh$,
is predicted to be lower than the one of the $Z$ boson. However, the
\onel\ corrections are known to be huge~\cite{mhiggs1l,mhoneloop}. 
As an impact,
$\mh > \MZ$ is possible, and an upper bound of
approximately $150 \gev$ is obtained. 
Hence a \twol\ calculation is inevitable for a precise 
prediction of the mass of the lightest Higgs boson. This is 
particularly important in view of the search for this particle at LEP2,
where a precise knowledge of $\mh$ in terms of the relevant SUSY
parameters is crucial in order to determine the discovery (and of course
also the exclusion) potential of LEP2.

Up to now there exist renormalization group improvements of the one-loop
result by including the \twol\ leading logarithmic
contributions~\cite{mhiggsRG1,mhiggsRG1a,mhiggsRG2}, and a
diagrammatic calculation of the
dominant \twol\ contributions in the limiting case of 
vanishing $\Stop$-mixing and infinitely large $\MA$ and 
$\Tb$~\cite{hoanghempfling}.
These results indicate that the \twol\ corrections considerably reduce
the prediction for $\mh$. However, a diagrammatic \twol\ calculation 
of the neutral mass spectrum
going beyond the above-mentioned limiting case
has been missing so far. 
Such a Feynman diagrammatic calculation is technically very involved,
but it is of particular interest since it allows for general
parameters of the MSSM Higgs sector and for virtual particle effects
without restrictions on their masses and mixing.
It is the purpose of this letter to
investigate the leading QCD corrections to the masses of the 
neutral $\cp$-even Higgs bosons and in particular to provide in this way 
a \twol\ prediction of $\mh$ for arbitrary values
of the parameters of the Higgs and scalar top sector of the MSSM.


Contrary to the SM, in the MSSM two Higgs doublets are needed.
The  Higgs potential is given by~\cite{hhg}
\BEA
\label{Higgspot}
V =& &m_1^2 H_1\bar{H}_1 + m_2^2 H_2\bar{H}_2 - m_{12}^2 (\epsilon_{ab}
      H_1^aH_2^b + \hc)  \nonumber \\
   &+&\frac{g'^2 + g^2}{8}\, (H_1\bar{H}_1 - H_2\bar{H}_2)^2
      +\frac{g^2}{2}\, |H_1\bar{H}_2|^2,
\EEA
where $m_1, m_2, m_{12}$ are soft SUSY-breaking terms, 
$g, g'$ are the $SU(2)$ and $U(1)$ gauge couplings, and 
$\epsilon_{12} = -1$.

The doublet fields $H_1$ and $H_2$ are decomposed  in the following way:
\BEA
H_1 &=& \VL H_1^1 \\ H_1^2 \VR = \VL v_1 + (\phi_1^{0} + i\chi_1^{0})
                                 /\sqrt2 \\ \phi_1^- \VR ,\non \\
H_2 &=& \VL H_2^1 \\ H_2^2 \VR =  \VL \phi_2^+ \\ v_2 + (\phi_2^0 
                                     + i\chi_2^0)/\sqrt2 \VR.
\label{eq:hidoubl}
\EEA
The potential \refeq{Higgspot} can be described with the help of two  
independent parameters (besides $g$, $g'$): 
$\Tb = v_2/v_1$ and $M_A^2 = -m_{12}^2(\Tb+\CTb)$,
where $M_A$ is the mass of the $\cp$-odd $A$ boson.



At tree level the mass matrix of the neutral $\cp$-even Higgs bosons
is given in the $\phi_1-\phi_2$ basis 
in terms of $\MZ$ and $\MA$ through
\BEA
&& M_{\rm Higgs}^{2, {\rm tree}} = \ML \mpe^2 & \mpez^2 \\ 
                           \mpez^2 & \mpz^2 \MR = \\
&& \ML \MA^2 \SQb + \MZ^2 \CQb & -(\MA^2 + \MZ^2) \Sb \Cb \\
    -(\MA^2 + \MZ^2) \Sb \Cb & \MA^2 \CQb + \MZ^2 \SQb \MR.\non
\EEA

The tree-level mass predictions receive large corrections at one-loop
order through terms proportional to $G_F \mt^4 \ln
(\mste \mstz/\mt^2)$~\cite{mhiggs1l}. These 
dominant one-loop contributions can be obtained by evaluating the
contribution of the $t-\Stop$-sector to the $\phi_{1,2}$ self-energies at zero
external momentum from the Yukawa part of the theory 
(neglecting the gauge couplings). Accordingly, the one-loop corrected 
Higgs masses are derived by diagonalizing the mass matrix
\BE
M^2_{\rm Higgs}
= \VL \mpe^2 - \hSi_{\Pe}(0)\;\;\;\;\;\; \mpez^2 - \hSi_{\PePz}(0) \\
     \mpez^2 - \hSi_{\PePz}(0)\;\;\;\;\;\; \mpz^2 - \hSi_{\Pz}(0) \VR ,
\label{higgsmassmatrixnondiag}
\EE
where the $\hSi$ denote the Yukawa contributions of the $t-\Stop$-sector
to the renormalized \onel\ $\phi_{1,2}$ self-energies. 
By comparison with the full \onel\ result~\cite{mhoneloop} it has been
shown that these contributions indeed contain the bulk of the \onel\
corrections. 
They typically approximate the full \onel\ result up to about 5~GeV.

In order to derive the leading two-loop contributions to the masses of
the neutral $\cp$-even Higgs bosons we have evaluated the QCD
corrections to \refeq{higgsmassmatrixnondiag}, which because of the
large value of the strong coupling constant are expected to be the most
sizable ones (see also \citere{hoanghempfling}). This requires the
evaluation of the renormalized $\phi_{1,2}$ self-energies at the
\twol\ level. Typical Feynman diagrams
corresponding to the Yukawa contributions of the $t-\Stop$-sector to the
$\phi_{1,2}$ self-energies and tadpoles are shown in \reffi{fig:fdtl}. They
have to be supplemented by the counterterm insertions in the
corresponding \onel\ diagrams.
Fig.~\ref{fig:fdtl}a shows the pure scalar
contributions to the Higgs self-energies. In Fig.~\ref{fig:fdtl}b the
gluonic corrections are depicted, while Fig.~\ref{fig:fdtl}c shows the
gluino-exchange contribution. In Fig.~\ref{fig:fdtl}d-f the tadpole
contributions for these three types of corrections are given.

\begin{figure}[tb]
\begin{center}
\mbox{
\psfig{figure=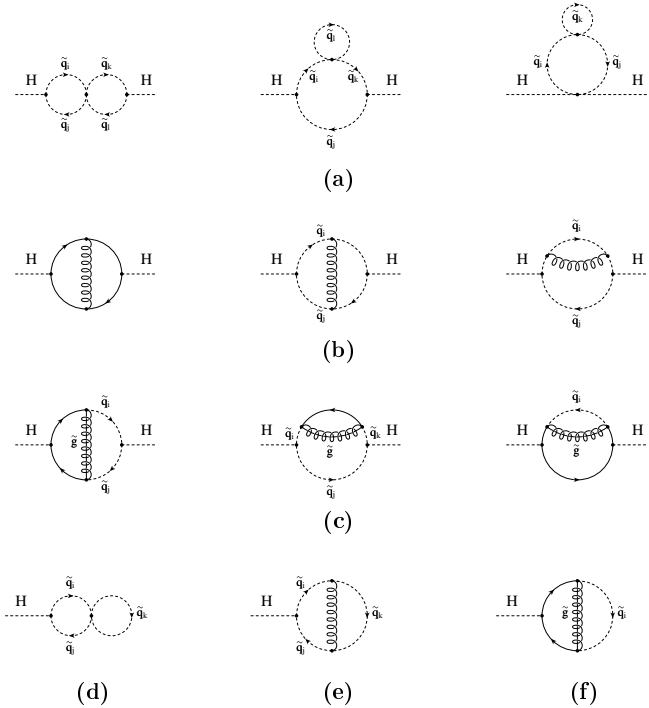,width=6.5cm,height=9.0cm,
                      bbllx=150pt,bblly=305pt,bburx=450pt,bbury=725pt}}
\end{center}
\caption[]{Typical Feynman diagrams for the two-loop contribution 
to the Higgs-boson self-energies and tadpoles. 
$H = \phi_1, \phi_2, A$.
}
\label{fig:fdtl}
\end{figure}

The renormalization has been performed in the on-shell scheme.
The counterterms in the Higgs sector are derived from the Higgs
potential~\refeqs{Higgspot}, (\ref{eq:hidoubl})
by expanding the counterterm contributions up
to \twol\ order. The renormalization conditions for
the tadpole counterterms have
been chosen in such a way that they cancel the tadpole contributions in
one- and \twol\ order. 
The renormalization in the $t-\Stop$-sector has
been performed in the same way as in \citere{drhosuqcd}.
For the present calculation the \onel\ counterterms $\delta m_t$,
$\delta m_{\Stope}$, $\delta m_{\Stopz}$ for the top-quark
and scalar top-quark masses and $\delta \tst$ for the mixing angle
contribute, which enter via the subloop renormalization. The appearance
of the 
$\Stop$ mixing angle $\tst$ reflects the fact that the current
eigenstates, $\StopL$ and $\StopR$, mix to give the mass eigenstates
$\Stope$ and $\Stopz$. 
Since the non-diagonal entry in the scalar
quark mass matrix is proportional to the
quark mass the mixing is particularly important in the case of the third
generation scalar quarks.
The mixing-angle counterterm
$\de \tst$ is chosen such that there is no
mixing between $\Stope$ and $\Stopz$ when $\Stope$ is
on-shell. The numerical result, however, is insensitive to this choice of the
renormalization point.
The one-loop counterterms for $\mu$ and $\Tb$, 
$\de\mu$ and $\de\Tb$, do not contribute since they are independent of
$\als$. 

The renormalized self-energies have the following structure:
\BE
\label{P1serenb}
\hSi_s(0) = \Sie_s(0) + \Siz_s(0)
- \de V_s^{(1)} - \de V_s^{(2)},
\EE
where $s = \phi_1, \phi_2, \phi_1 \phi_2$. $\Sie_s$ and $\Siz_s$ denote
the unrenormalized self-energies at the one- and \twol\ level, and
$\de V^{(1)}_s$ and $\de V^{(2)}_s$ are the one- and \twol\ counterterms 
derived from the Higgs potential. The counterterms 
read:
\BEA
\label{P1potctMW0Yuk1l}
\de V_{\Pe}^{(i)} &=& 
             + \de\MA^{2(i)} \SQb
             - \de t_1^{(i)} \frac{e\, \Cb}{2 \MW \sw} (1 + \SQb) \non\\
         &&  + \de t_2^{(i)} \frac{e}{2 \MW \sw} \CQb \Sb, \\
\label{P2potctMW0Yuk1l}
\de V_{\Pz}^{(i)} &=& 
             + \de\MA^{2(i)} \CQb
             - \de t_2^{(i)} \frac{e\, \Sb}{2 \MW \sw} (1 + \CQb) \non\\
         &&  + \de t_1^{(i)} \frac{e}{2 \MW \sw} \SQb \Cb, \\
\label{P1P2potctMW0Yuk1l}
\de V_{\PePz}^{(i)} &=& 
               - \de\MA^{2(i)} \Sb\Cb
               - \de t_1^{(i)} \frac{e}{2 \MW \sw} \SDb \non\\
         &&    - \de t_2^{(i)} \frac{e}{2 \MW \sw} \CDb, 
\EEA
with $\de t_a^{(i)} = - T_a^{(i)}$, where $T_a^{(i)}$ denotes the tadpole
contribution, $\de t_a^{(i)}$ is the corresponding counterterm, 
and $\de\MA^{2(i)} = \Si_A^{(i)}(0)$ ($i = 1,2$).

In deriving our results we have made strong use of computer-algebra
tools. 
The package \fa~\cite{feynarts} (in which the relevant part
of the MSSM has been implemented) has been applied to generate the
Feynman amplitudes and the counterterm contributions. For evaluating the
amplitudes the package \tc~\cite{twocalc} has been used.
The calculations have been performed using Dimensional Reduction
(DRED)~\cite{dred}, which is necessary in order to preserve the
relevant SUSY relations. Naive application (without an appropriate 
shift in the couplings) of Dimensional Regularization
(DREG)~\cite{dreg}, on the other hand, does not lead to a finite result.
The same observation has also been made in~\citere{hoanghempfling}.

The contributions of the scalar, the gluon-, and the gluino-exchange
diagrams in \reffi{fig:fdtl} together with the corresponding counterterm
contributions are not separately finite (as it was the case in the
calculation of \citere{drhosuqcd}), but have to be combined in
order to obtain a finite result. Our results for the \twol\
$\phi_{1,2}$ self-energies are given in terms of the
SUSY parameters $\Tb$, $\MA$, $\mu$, $\mste$, $\mstz$, $\tst$, and 
$\mgl$. In the general case the results
are by far too lengthy to be given here
explicitly. In the special case of vanishing mixing in the 
$\Stop$-sector, $\mu = 0$, and $\mste = \mstz = \mst$, a relatively 
compact expression can be derived. It is given by
\BE
\hSiz_{\Pe}(0) = 0, \;\;\;\;\; \hSiz_{\PePz}(0) = 0
\EE
\BEA
\lefteqn{
\hSiz_{\Pz}(0) = \frac{G_F \sqrt{2}}{\pi^2} \frac{\als}{\pi}
\frac{\mt^2}{\SQb} \Bigg\{ \KL \mst^2 - \mgl^2 - \mt^2 \KR\cdot } \non \\
&& \Big[ (1 + \frac{\mt^2}{\mst^2} ) 
 - \Big( \re B_0^{fin}(\mt^2, \mgl, \mst) \KL 1 - 2 L \KR \non \\
&& {}+ \re B_0^{fin}(\mst^2, \mgl, \mt) \frac{\mt^2}{\mst^2} \Big) \non\\ 
&& {}- \frac{\mgl^2}{\mst^2\,N} 
        \KL \mgl^2 (\mst^2 + \mt^2) - (\mst^2 - \mt^2)^2 \KR 
        \ln \KL \mgl^2 \KR \non \\
&& {} + \frac{4}{N^2} \mgl^4 \mt^2 \Phi(\mt, \mst, \mgl) \Big]
   - 2 \mgl^2 \ln\KL \mgl^2 \KR L \non \\
&& {} + (2 \mst^2 + \mt^2) \ln \KL \mst^2 \KR L 
      - 3 \mt^2 \ln \KL\mt^2 \KR \ln \KL\mst^2\KR \non \\
&& {} + \frac{1}{N} \ln \KL \mst^2 \KR \Big[ 2 \mgl^6 
      - \mgl^4 (7 \mst^2 + \mt^2) + 4 \mgl^2 (2 \mst^4 \non \\
&& {} - 3 \mst^2 \mt^2 - 3 \mt^4) 
      -(3 \mst^2 - 7 \mt^2) (\mst^2 - \mt^2)^2 \Big]\non \\
&& {} + 3 \mt^2 \ln^2 \KL \mt^2 \KR 
      + \frac{1}{\mst^2 N} \ln \KL \mt^2 \KR 
        \Big[ 2 \mst^2 (\mst^2 - \mgl^2)^3\non \\
&& {} - \mst^2 \mt^2 (5 \mgl^4 - 16 \mgl^2 \mst^2 + 11 \mst^4) 
      + \mt^4 (17 \mst^4 \non \\
&& {} + 6 \mgl^2 \mst^2 - \mgl^4) 
      - 9 \mst^2 \mt^6 + \mt^8 \Big] \Bigg\} , 
\label{eq:resphi2spec} 
\EEA
with $L = \ln(\mst^2/\mt^2)$, and
\BEA
&& N = \KL (\mgl - \mt)^2 - \mst^2 \KR 
       \KL (\mgl + \mt)^2 - \mst^2 \KR, \non \\
&& \Phi(x,y,z) = \edz z^2 \la \KL \frac{x^2}{z^2}, \frac{y^2}{z^2} \KR
   \Big[ 2 \ln(\al^1_{xyz}) \ln(\al^2_{xyz}) \non \\
&& {}    - \ln \KL \frac{x^2}{z^2} \KR \ln \KL \frac{y^2}{z^2} \KR
         - 2 {\rm Li_2}(\al^1_{xyz}) - 2 {\rm Li_2}(\al^2_{xyz})
         + \frac{\pi^2}{3} \Big], \non \\
&& \la(u, v) = \sqrt{1 + u^2 + v^2 - 2 u - 2 v - 2 u v}, \non \\
&& \al^j_{xyz} = \edz\Big[1 - (-1)^j \frac{x^2}{z^2} 
                            + (-1)^j \frac{y^2}{z^2}
    - \la\KL \frac{x^2}{z^2} ,\frac{y^2}{z^2} \KR \Big], \non \\
&& B_0^{fin}(p^2,m_1,m_2) = -\ln(m_1) -\ln(m_2) + 2 \non \\
&& {} - \frac{m_1^2/m_2^2 - 1}{2 p^2/m_2^2} 
     \ln (m_1^2/m_2^2) 
   + \frac{r_1 - r_2}{2 p^2/m_2^2} \KL\ln(r_1) - \ln(r_2)\KR , \non
\EEA
$r_j$ being the solutions of
$m_2^2 r + m_1^2/r = m_1^2 + m_2^2 - p^2$ ($j = 1,2$). 
Eq.~(\ref{eq:resphi2spec}) approximates the complete numerical
result for vanishing mixing (for arbitrary $\mu$ and $\mste \neq \mstz$)
up to about $2\%$ accuracy.

Inserting the \onel\ and \twol\ $\phi_{1,2}$ self-energies into 
\refeq{higgsmassmatrixnondiag}, the predictions for the masses of the
neutral $\cp$-even Higgs bosons are derived by diagonalizing the 
\twol\ mass matrix.
For the numerical evaluation we have chosen two values for $\Tb$ which
are favored by SUSY-GUT scenarios~\cite{su5so10}: $\Tb = 1.6$ for
the $SU(5)$ scenario and $\Tb = 40$ for the $SO(10)$ scenario. Other
parameters are $\MZ = 91.187 \gev, \MW = 80.375 \gev, 
G_F = 1.16639 \, 10^{-5} \gev^{-2}, \als = 0.1095$, and $\mt = 175 \gev$. 
For the figures below we have furthermore
chosen $\mu = - 200 \gev, \MA = 500 \gev$, and $\mgl = 500 \gev$ as typical
values. The scalar top masses and the mixing angle are derived from the
parameters $M_{{\tilde t}_L}$, $M_{{\tilde t}_R}$ and $\Mtlr$ of the 
$\Stop$ mass matrix (our conventions are the same as in
\citere{drhosuqcd}). In the figures below we have chosen 
$\msq \equiv \MstL = \MstR$.

The plot in Fig.~\ref{fig:plot1} shows $\mh$
as a function of $\Mtlr/\msq$, where $\msq$
is fixed to $500 \gev$. A minimum is reached for  
$\Mtlr = 0 \gev$ which we refer to as `no mixing'. A maximum in the
\twol\ result for $\mh$ is reached for about $\Mtlr/\msq \approx
2$ in the $\Tb = 1.6$ scenario as well as in the $\Tb = 40$ scenario. 
This case we refer to as `maximal mixing'. 
Note that the maximum is shifted compared to its \onel\ value of about
$\Mtlr/\msq \approx 2.4$.

\begin{figure}[htb]
\begin{center}
\mbox{
\psfig{figure=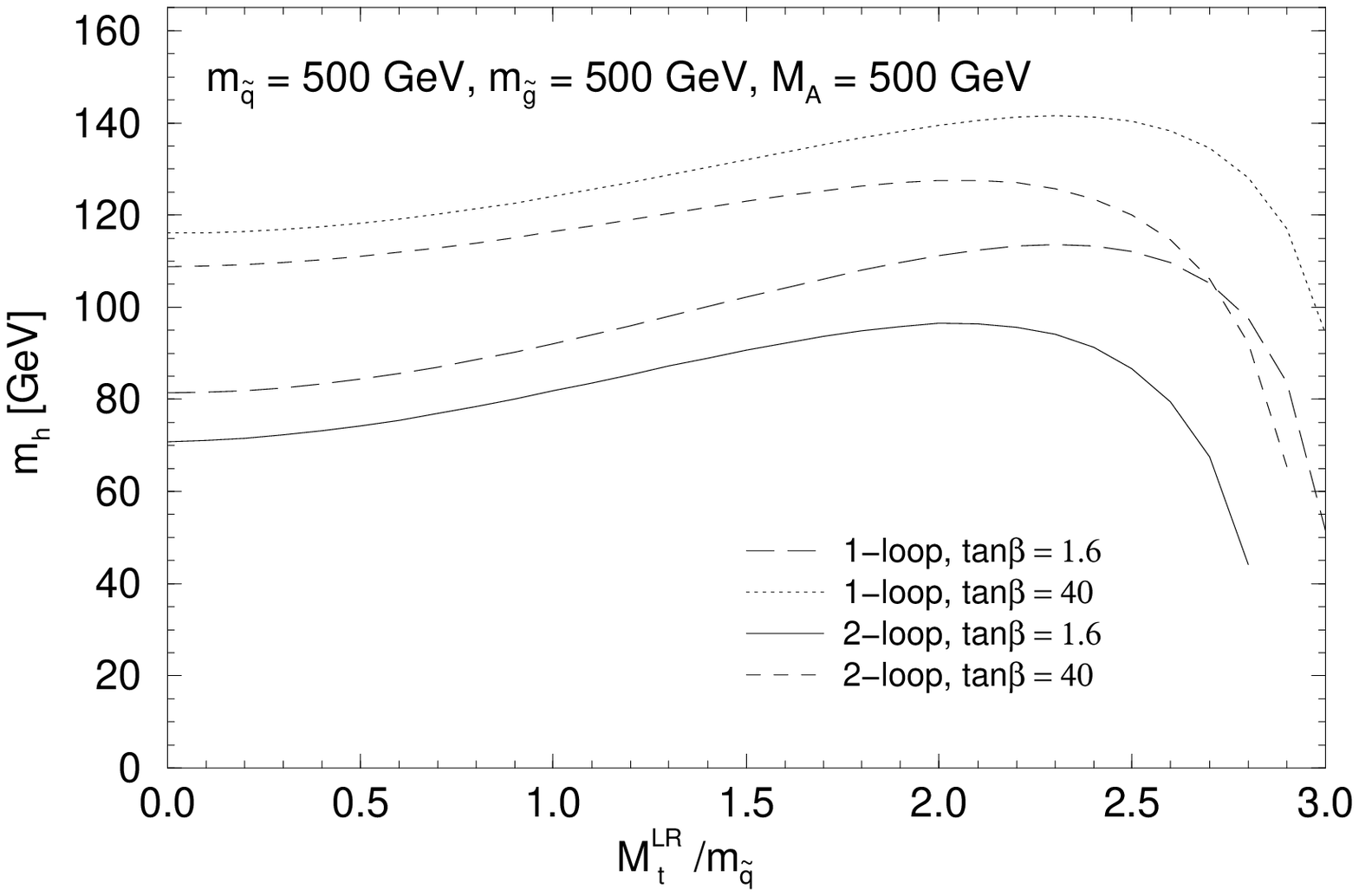,width=5.3cm,height=8cm,
                      bbllx=150pt,bblly=100pt,bburx=450pt,bbury=420pt}}
\end{center}
\caption[]{One- and \twol\ results for $\mh$ as a function of
$\Mtlr/\msq$ for two values of $\Tb$.}
\label{fig:plot1}
\end{figure}

In Fig.~\ref{fig:plot2} the low-$\Tb$ scenario with $\Tb = 1.6$ is
analyzed. The tree-level, the \onel\ and the \twol\ results for $\mh$
are shown as a function of $\msq$ for no mixing and maximal mixing. 
For both cases the \onel\ result is in general considerably reduced.
For the no-mixing
case the difference between the \onel\ and \twol\ result amounts up to
about $18 \gev$ for $\msq = 1 \tev$.
In the maximal-mixing case the reduction of the \onel\ result 
is about $10 \gev$ for $\msq = 260 \gev$ (for smaller $\msq$
one gets unphysical or experimentally excluded $\Stop$-masses) and
about $25 \gev$ for $\msq = 1 \tev$. 

\begin{figure}[htb]
\begin{center}
\mbox{
\psfig{figure=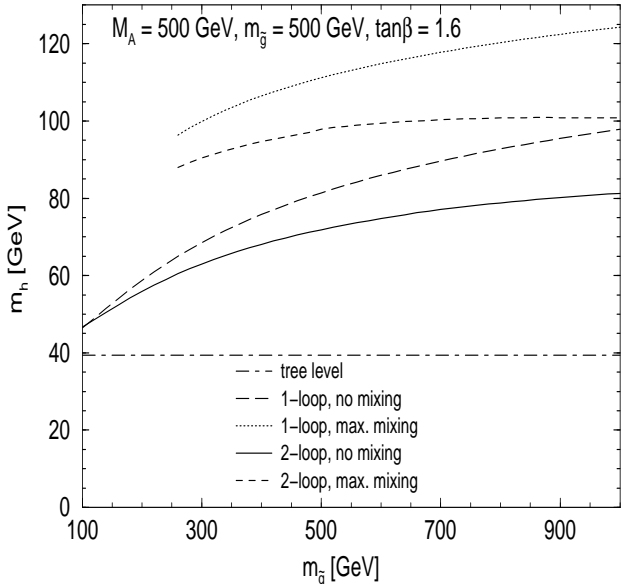,width=5.3cm,height=7.8cm,
                      bbllx=150pt,bblly=100pt,bburx=450pt,bbury=410pt}}
\end{center}
\caption[]{
The mass of the lightest Higgs boson for $\Tb = 1.6$. 
The tree-, the one- and the \twol\ results for $\mh$ are shown 
as a function of $\msq$ for the no-mixing and the maximal-mixing case.}
\label{fig:plot2}
\end{figure}

The variation of this 
result with $\mgl$ is of the order of few GeV. Varying $\Tb$ 
around the value $\Tb = 1.6$ leads to a relatively large effect in 
$\mh$. Higher values for $\mh$ are obtained for larger
$\Tb$. A more detailed analysis of the dependence of our results on the 
different SUSY parameters will be presented in a forthcoming publication. 

In Fig.~\ref{fig:plot3} the high-$\Tb$ scenario with $\Tb = 40$ is
analyzed. Again the tree-level, the \onel\ and the \twol\ results for
$\mh$ are shown as a function of $\msq$ for minimal and maximal mixing. 
As in the case of low $\tan\beta$, the \onel\ result is in general
considerably reduced. 
For no mixing the difference between the \onel\ and \twol\ result reaches
about $14 \gev$ for $\msq = 1 \tev$. 
In the maximal-mixing case the
reduction of the \onel\ result amounts to about $7 \gev$ for 
$\msq = 260 \gev$ and about $22 \gev$ for $\msq = 1 \tev$.
The reduction of the \onel\ result is slightly smaller than for $\Tb = 1.6$. 
This can be understood from the result for $\hSi_{\Pz}(0)$ given as a
special case in \refeq{eq:resphi2spec}.
In this 
case $\be$ appears only in the prefactor as $1/\SQb$ and 
one thus gets a bigger reduction of $\mh$ for smaller $\tan \beta$.
The variation of the result shown in \reffi{fig:plot3} with
$\mgl$ is again of the order of few GeV. The effect of varying 
$\Tb$ around $\Tb = 40$ is marginal.

\begin{figure}[htb]
\begin{center}
\mbox{
\psfig{figure=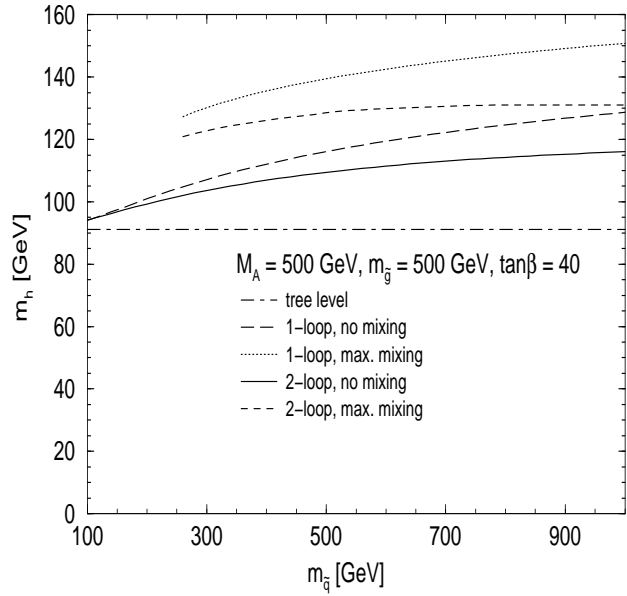,width=5.3cm,height=8cm,
                      bbllx=150pt,bblly=100pt,bburx=450pt,bbury=420pt}}
\end{center}
\caption[]{
The mass of the lightest Higgs boson for $\Tb = 40$. 
The tree-, the one- and the \twol\ results for $\mh$ are shown
as a function of
$\msq$ for the no-mixing and the maximal-mixing case.}
\label{fig:plot3}
\end{figure}


We have compared our results with the results obtained in
\citere{hoanghempfling} in the case of no $\Stop$-mixing and $\MA \to \infty,
\Tb \to \infty$ and have checked analytically that in the limiting case
$\mste = \mstz = \mgl \gg \mt$ 
in \refeq{eq:resphi2spec} 
we recover the corresponding formula
given in \citere{hoanghempfling}.

Supplementing our results for the leading $\oaas$ corrections with the
leading higher-order Yukawa term of ${\cal O}(\alpha^2 \mt^6)$ given in
\citere{mhiggsRG1a} leads to an increase in the prediction of $\mh$ of
up to about 3~GeV.
A similar shift towards higher values of $\mh$ emerges if at
the two-loop level the running top-quark mass, 
${\overline\mt}(\mt) = 166.5 \gev$, is
used instead of the pole mass, $\mt = 175 \gev$, thus taking into
account leading higher-order effects beyond the \twol\ level. 
We have compared our results with
the results obtained by
\twol\ renormalization group calculations 
given in \citeres{mhiggsRG1,mhiggsRG2}.%
\footnote{
The results of \citere{mhiggsRG1} and \citere{mhiggsRG2} agree within about
$2 \gev$ with each other.
}
We find good 
agreement for the case of no $\Stop$-mixing, while for larger
$\Stop$-mixing sizable deviations exceeding 5~GeV occur. In
particular, the value of $\Mtlr/\msq$ yielding the maximal $\mh$
is shifted from $\Mtlr/\msq \approx 2.4$ in the \onel\ case to 
$\Mtlr/\msq \approx 2$ when our
diagrammatic \twol\ results are included (see \reffi{fig:plot1}).
In the results based on renormalization group 
methods~\cite{mhiggsRG1,mhiggsRG2}, on the other hand, the maximal
value of $\mh$ is obtained for $\Mtlr/\msq \approx 2.4$, i.e.\ at the
same value as for the \onel\ result.

In summary, we have diagrammatically calculated the leading $\oaas$
corrections to the masses of the neutral $\cp$-even Higgs bosons in 
the MSSM. 
We have applied the on-shell scheme and have imposed no restrictions on
the parameters of the Higgs and scalar top sector of the model. 
The \twol\ correction leads to a considerable reduction of the
prediction for the mass of the lightest Higgs boson compared to the 
\onel\ value. 
The reduction turns out to be particularly important for
low values of $\Tb$.
The leading \twol\ contributions presented here can directly be combined
with the complete \onel\ results in the on-shell scheme~\cite{mhoneloop}.
A discussion of the corresponding results will be given in a
forthcoming paper, where also a more detailed comparison with the
results based on renormalization group methods will be pursued.\\


We thank M.~Carena, H.~Haber and C.~Wagner for fruitful discussions and 
communication about the numerical comparison of our results. We also 
thank A.~Djouadi and H.~Eberl for valuable discussions.


\end{narrowtext}
\end{document}